\documentclass[12pt]{iopart}
 \usepackage{graphicx,amsfonts,color,amssymb,latexsym,amsbsy}

 \newcommand\Rb{\mathbb{R}}

 \newcommand{\ben}{\begin{equation*}}
 \newcommand{\ebn}{\end{equation*}}
 \newcommand\be{\begin{equation}}
 \newcommand\eb{\end{equation}}
 
 \newcommand\Kr{\mathcal{K}}

\begin{document}
 \title{Form factors of twist fields in the lattice Dirac theory}
 \author{P Gavrylenko$^1$, N Iorgov$^2$ and O Lisovyy$^3$}
 \address{
  $^1$ Department of Physics, Kyiv National University, 03022 Kyiv, Ukraine}
 \address{
  $^2$ Bogolyubov Institute for Theoretical Physics,
 03680 Kyiv, Ukraine}
  \address{$^3$ Laboratoire de Math\'ematiques et Physique Th\'eorique CNRS/UMR 6083,
  Universit\'e de Tours, Parc de Grandmont, 37200 Tours, France}
  \eads{\mailto{iorgov@bitp.kiev.ua}, \mailto{lisovyi@lmpt.univ-tours.fr}}

 \begin{abstract}
 We study $U(1)$ twist fields in a two-dimensional lattice theory of massive
 Dirac fermions. Factorized formulas for finite-lattice
 form factors of these fields are derived using elliptic parametrization of the spectral curve
 of the model, elliptic determinant identities and theta functional interpolation.
 We also investigate the thermodynamic and the infinite-volume scaling limit,
 where the corresponding expressions reduce to form factors of the exponential fields
 of the sine-Gordon model at the free-fermion point.
 \end{abstract}


 \section{Introduction}
 It is a general property of two-dimensional quantum field theories that their
 symmetries give rise to new local fields, whose correlation functions
 are nontrivial even if the underlying theory is free. Two paradigmatic examples are given
 by the disorder variables in the Ising field theory \cite{Kadanoff,Truong} and $U(1)$ twist fields in the
 massive Dirac theory \cite{Marino,smj,Truong2}, directly related
 to the exponential fields in the sine-Gordon model at the free-fermion point.

 Correlation functions of twist fields in the Dirac theory, as well as in its generalizations
 to curved space and non-zero background magnetic field
 \cite{Doyon,Doyon_Fonseca,lisovyy_JMP,PBT}, are also interesting from the mathematical point
 of view. They satisfy
 nonlinear differential equations \cite{Bernard_Leclair,Doyon_Silk,smj},
  which in the simplest cases can be solved in terms of Painlev\'e functions.
  The knowledge of the long- and short-distance behaviour of the two-point correlators
  provides solutions to nontrivial asymptotic and connection problems of Painlev\'e theory.
 Recently, it has also been observed \cite{lisovyy_dyson} that such correlators coincide with the gap probabilities for the classical kernels arising in the representation theory of big groups \cite{borodin}.

 The aim of this paper is to construct lattice analogs of $U(1)$ twist fields in the Dirac model,
 satisfying the following properties:
 (i) they should be defined via the branching of lattice fermion fields,
 (ii) one should be able to calculate their form factors explicitly and
 (iii) these form factors should reproduce the known expressions in the scaling limit.
 Besides full control of the theory, such an integrable finite-lattice regularization
 can be used for investigations at non-zero temperature and
 for a mathematically sound derivation of the relative normalization of
 conformal and infrared asymptotics of the two-point correlator \cite{Bugrij_Shadura,LZ}.
 It may also be instrumental in going beyond the free-fermion point.

 While the Ising field theory possesses a natural lattice regularization, only
 a few results are available in the Dirac case. First attempt to introduce
 twist fields on the infinite lattice was made in \cite{Palmer_MF}. The corresponding definition
 was supported by the computation of the  vacuum
 expectation value (reproducing the expected scaling dimension),
 and was further strengthened by the analysis of correlations at the critical point \cite{Palmer_SD}.
 Another, seemingly unrelated definition was used in \cite{Bugrij_Shadura,lisovyy_peyresq} to derive a number of  determinant representations for the two-point function of lattice twist fields.
 The present work is devoted to the computation of their form factors, i.~e. matrix elements
 of the field operators in the basis of transfer matrix eigenstates.

 The paper is planned as follows. In Subsections~\ref{sec21} and~\ref{sec22}, we introduce a one-parameter generalization of the lattice Dirac operator considered in \cite{Bugrij_Shadura,lisovyy_peyresq}
 and explain the definition of twist fields in the functional integral framework. Transition to the
 operator formalism is performed in the next subsection. Using the coherent states approach, the transfer matrix and the twist field operator are written as exponentials of fermion bilinears, see formulas (\ref{tm})
 and (\ref{twf}) below. Subsection~\ref{sec24} is devoted to the construction of multiparticle
 Fock states simultaneously diagonalizing the transfer matrix and the operator of translations.

 In Subsection~\ref{sec31}, it is explained how form factors of twist fields can be found
 from the linear transformations relating fermionic creation-annihilation operators of
 different periodicity. In particular, the vacuum expectation value and two-particle form factors
 are expressed in terms of two square matrices $C$ and $D$ of dimension equal to the lattice size.
 Essentially, one needs to compute the quantities $D^{-1}$, $D^{-1}C$ and $\mathrm{det}\,D$, cf.~(\ref{ffvev})--(\ref{dchi}). This task
 is solved in Subsection~\ref{sec32} by first noting that in the elliptic parametrization of the
 spectral curve of the model $C$ and $D$ are given (up to diagonal factors) by elliptic Cauchy matrices,
 and then using Frobenius
 determinant identity and theta functional interpolation along the lines of \cite{IL}.
 The corresponding expressions are further simplified in Subsection~\ref{sec33}. Finite-lattice
 two-particle form factors are given by (\ref{dmonec2}), (\ref{detd2}) and (\ref{dmone2}), and the
 multiparticle ones have the factorized form (\ref{muff}), (\ref{detr}). These formulas
 represent the main result of the paper. In Subsection~\ref{sec34}
 we analyze the thermodynamic (infinite-lattice) limit. The final answer
 has a remarkably simple expression in terms of the Jacobi theta functions, see
 (\ref{vevfinal})--(\ref{ffncth}). We remark that the vacuum expectation
  value (\ref{vevfinal}) reproduces the earlier result of \cite{Palmer_MF}.
  Field theory limit is considered in Subsection~\ref{sec35}. It is shown
 that the scaled form factors coincide with those of the exponential fields of sine-Gordon model at
 the free-fermion point \cite{Bernard_Leclair,Marino,Truong2}. We conclude with a brief discussion
 of results and open problems.

 \section{Lattice Dirac theory}
 \subsection{Fermions}\label{sec21}
 Let $\psi$, $\bar{\psi}$ denote two 2-component Grassmann fields  on an $M\times N$
 square lattice. Consider the standard free-fermion action $S[\psi,\bar{\psi}]=\bar{\psi}D\psi$,
 where the lattice Dirac operator is chosen as
 \be\label{dirop}
 D=\frac{1}{c_x^*}\left(\begin{array}{cc}
 c_x^*s_y-s_x^*c_y\nabla_y & -c_y+c_x^*\nabla_{x} \\ c_y-c_x^*\nabla_{-x} &
 c_x^*s_y-s_x^*c_y\nabla_{-y}
 \end{array}\right).
 \eb
 Here $\nabla_{x,y}$ denote the shifts by one lattice site in the horizontal and vertical directions,
 so that e.~g. $\nabla_{x}\psi_{x,y}=\psi_{x+1,y}$, $\nabla_{y}\psi_{x,y}=\psi_{x,y+1}$.
 The boundary conditions with respect to $x$ and $y$ are antiperiodic and $\alpha$-periodic,
 respectively. This means
 that
 \ben
 \cases{\psi_{x+M,y}=-\psi_{x,y},\\\bar{\psi}_{x+M,y}=-\bar{\psi}_{x,y},}\qquad
 \cases{\psi_{x,y+N}=e^{2\pi i \alpha}\psi_{x,y},\\\bar{\psi}_{x,y+N}=e^{-2\pi i \alpha}\bar{\psi}_{x,y}.}
 \ebn
 The parameters $c_i$, $s_i$, $c_i^*$, $s_i^*$ ($i=x,y$) are expressed in terms of two constants
 $\Kr_{x,y}\in\mathbb{R}_{>0}$ as
 \ben
 c_i=\cosh2\Kr_i,\quad s_i=\sinh2\Kr_i,\quad
 c^*_i=\cosh2\Kr^*_i,\quad s^*_i=\sinh2\Kr_i^*,
 \ebn
 where  the dual couplings are given by $\Kr^*_i=\mathrm{arctanh}\,e^{-2\Kr_i}$. They
 satisfy  Ising-type relations $s_i^*=s_i^{-1}$, $c_i^*=c_is_i^{-1}$. It will be assumed in the
 following that $\Kr_x^*<\Kr_y$. Dirac operator considered in \cite{Bugrij_Shadura,lisovyy_peyresq}
 is obtained from  (\ref{dirop}) by setting $\mathcal{K}_x=\mathcal{K}_y$.

 Fermion propagator can be found using Fourier transform. One obtains
 \be\label{fprop}
 \fl\langle \bar{\psi}_{x,y}\psi_{x',y'}\rangle=\frac{1}{MN}\sum_{\phi,\theta}
 \frac{ \,e^{i\phi(x-x')+i\theta(y-y')}}{
 2c_y\left(\cosh\gamma_{\theta}-\cos \phi\right)}
 \left(\begin{array}{cc}
 c_x^*s_y-s_x^*c_y e^{-i\theta} & c_y-c_x^*e^{i\phi} \\ -c_y+c_x^*e^{-i\phi} & c_x^*s_y-s_x^*c_y e^{i\theta}\end{array}\right),
 \eb
 where
 \be\label{gammatheta}
 \cosh\gamma_{\theta}=c_x^*c_y-s_x^*s_y\cos\theta.
 \eb
 The multipoint correlations are readily computable from the Wick theorem. The summation
 in (\ref{fprop}) is performed over $\phi=\frac{2\pi}{M}\left(j+\frac12\right)$, $j=0,\ldots,M-1$ and
 $\theta=\frac{2\pi}{N}\left(k+\alpha\right)$, $k=0,\ldots,N-1$. In the thermodynamic limit
 $M,N\rightarrow\infty$ one has $\frac{1}{MN}\sum\limits_{\phi,\theta}\rightarrow \frac{1}{(2\pi)^2}\int
 \!\!\int\nolimits_0^{2\pi}d\phi\,d\theta$ and the answer can be expressed in terms of elliptic
 integrals.

 \subsection{Twist fields}\label{sec22}
 It is instructive to start with an example.
 Choose a closed path $\mathcal{P}$
 on the dual lattice (Fig.~1) and
 make the transformation $\psi\mapsto e^{2\pi i \nu}\psi$, $\bar{\psi}\mapsto e^{-2\pi i \nu}\bar{\psi}$
 with $\nu\in\Rb$ at all lattice sites inside this contour. Because of the global
 $U(1)$-symmetry, the action $S[\psi,\bar{\psi}]$ will change only at the edges intersected by $\mathcal{P}$.

 \begin{figure}[!h]
 \begin{center}
 \resizebox{8cm}{!}{
 \includegraphics{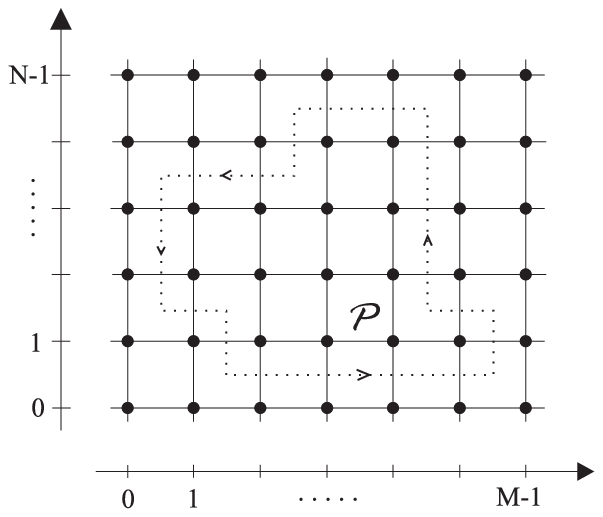}} \\
 Fig. 1
 \end{center}
 \end{figure}

 If the corresponding changes are made along an open path $\mathcal{P}_{AB}$
 joining two points $A$ and $B$ on the dual lattice, the resulting functional integral
 will depend on the positions of these points and homotopy class of the path, but
 not on its shape.
  One way to choose $\mathcal{P}_{AB}$ is shown in Fig.~2. In this case, the action is modified
 by
 \be\label{dSAB}
 \delta S_{AB}=\delta S_A+\delta S_B+\delta S_{\mathrm{b.c.}},
 \eb
 where $\delta S_{A,B}$ correspond to the vertical segments and $\delta S_{\mathrm{b.c.}}$
 to the horizontal one. Explicitly,
 \begin{eqnarray*}
 \delta S_A=2i\sin\pi\nu\sum_{y''=0}^{y-1}\left(e^{i\pi \nu}\bar{\psi}^1_{x-1,y''}\psi^2_{x,y''}+
 e^{-i\pi \nu}\bar{\psi}^2_{x,y''}\psi^1_{x-1,y''}\right),\\
 \delta S_B=-2i\sin\pi\nu\sum_{y''=0}^{y'-1}\left(e^{-i\pi \nu}\bar{\psi}^1_{x'-1,y''}\psi^2_{x',y''}+
 e^{i\pi \nu}\bar{\psi}^2_{x',y''}\psi^1_{x'-1,y''}\right),\\
 \fl\qquad\delta S_{\mathrm{b.c.}}=-2i\sin\pi\nu\,\frac{c_y}{c_x}\sum_{x''=x}^{x'-1}\left(
 e^{i\pi (\alpha+\alpha')}\bar{\psi}^1_{x'',N-1}\psi^1_{x'',0}-e^{-i\pi (\alpha+\alpha')}\bar{\psi}^2_{x'',0}\psi^2_{x'',N-1}\right),
 \end{eqnarray*}
 with  $\alpha'=\alpha+\nu$. Without any loss of generality, we assume that
 $0\leq \alpha,\alpha'<1$.

 \begin{figure}[!h]
 \begin{center}
 \resizebox{8cm}{!}{
 \includegraphics{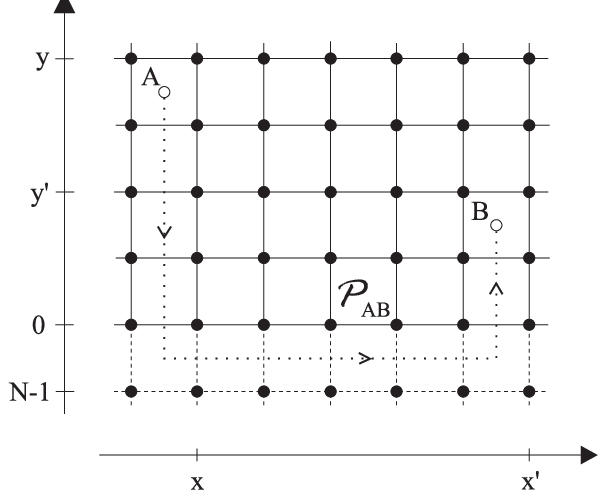}} \\
 Fig. 2
 \end{center}
 \end{figure}

 Twist fields live on the dual lattice. Their two-point correlation function is defined as
 the normalized partition function of the Dirac theory with the defect contribution (\ref{dSAB}):
 \be\label{2pcf}
 \left\langle\mathcal{O}_{\alpha,\alpha'}(A)\mathcal{O}_{\alpha',\alpha}(B)\right\rangle=
 \frac{\int\mathcal{D}\psi\mathcal{D}\bar{\psi}\;e^{S[\psi,\bar{\psi}]+\delta S_{AB}}}{\int\mathcal{D}\psi\mathcal{D}\bar{\psi}\;e^{S[\psi,\bar{\psi}]}}.
 \eb
 Note that the effect of $\delta S_{\mathrm{b.c.}}$ amounts to the change of the vertical boundary conditions
 for fermions from the horizontal interval $[x,x'-1]$ from $\alpha$- to $\alpha'$-periodic ones.
 More generally, the appearance of twist field $\mathcal{O}_{\alpha,\alpha'}(A)$ in an
 arbitrary correlation
 function means that
 \begin{itemize}
 \item the vertical boundary conditions for fermions to the left and right of $A$ are $\alpha$- and
 $\alpha'$-periodic, respectively;
 \item the term $\delta S_A$ should be added to the action.
 \end{itemize}
 In the heuristic continuum limit, this corresponds to
 integrating over field configurations having counterclockwise monodromy $e^{2\pi i\nu}$ of $\psi$
 (resp. $e^{-2\pi i\nu}$ for $\bar{\psi}$) around $A$.

 \subsection{Coherent states and the operator formalism}
 Locality of twist fields becomes manifest in the operator formalism, which also provides
 a convenient framework for the computation of correlation functions.

 Let us introduce two sets of fermionic creation-annihilation operators satisfying canonical anticommutation relations
 \ben
 \{a_y,a^{\dag}_{y'}\}=\{b_y,b^{\dag}_{y'}\}=\delta_{yy'},\qquad y,y'=0,\ldots,N-1,
 \ebn
 with all other anticommutators vanishing. Define in the usual way the vacuum vectors $\langle vac|$
 and $|vac\rangle$, normalized as $\langle vac|vac\rangle=1$, and the corresponding $2^{2N}$-dimensional
 Fock space $\mathcal{F}$. Further, introduce the coherent states
 \begin{eqnarray*}
 |\boldsymbol{\psi}^1_x,\bar{\boldsymbol{\psi}}_x^1\rangle=&\,\exp{\sum_{y=0}^{N-1}
 \left(a^{\dag}_y\bar{\psi}^1_{x,y}+b^{\dag}_y{\psi}^1_{x,y}\right)}|vac\rangle,\\
 \langle\boldsymbol{\psi}^2_x,\bar{\boldsymbol{\psi}}^2_x|=&\,\langle vac|
 \exp{\sum_{y=0}^{N-1}
 \left(\psi^2_{x,y}a_y+\bar{\psi}^2_{x,y}b_y\right)},
 \end{eqnarray*}
 where $\{\psi^i_{x,y}\}$,  $\{\bar{\psi}^i_{x,y}\}$ denote Grassmann variables anticommuting
 with all creation-annihilation operators. These states satisfy the following standard properties:
 \begin{itemize}
 \item For $y=0,\ldots,N-1$ one has
 \begin{eqnarray*}
 \cases{
 a_y|\boldsymbol{\psi}^1_x,\bar{\boldsymbol{\psi}}_x^1\rangle=
 \bar{\psi}^1_{x,y}|\boldsymbol{\psi}^1_x,\bar{\boldsymbol{\psi}}_x^1\rangle, \\
 b_y|\boldsymbol{\psi}^1_x,\bar{\boldsymbol{\psi}}_x^1\rangle=
 {\psi}^1_{x,y}|\boldsymbol{\psi}^1_x,\bar{\boldsymbol{\psi}}_x^1\rangle,
 }
 \\
 \cases{
 \langle\boldsymbol{\psi}^2_x,\bar{\boldsymbol{\psi}}^2_x|a^{\dag}_y=
 \langle\boldsymbol{\psi}^2_x,\bar{\boldsymbol{\psi}}^2_x|\psi^2_{x,y},\\
  \langle\boldsymbol{\psi}^2_x,\bar{\boldsymbol{\psi}}^2_x|b^{\dag}_y\,=
 \langle\boldsymbol{\psi}^2_x,\bar{\boldsymbol{\psi}}^2_x|\bar{\psi}^2_{x,y}.
 }
 \end{eqnarray*}
 \item The scalar product of two coherent states is given by
 \be\label{mecst}
 \langle\boldsymbol{\psi}^2_{x'},\bar{\boldsymbol{\psi}}^2_{x'}|
 \boldsymbol{\psi}^1_{x},\bar{\boldsymbol{\psi}}_{x}^1\rangle
 =\exp\sum_{y=0}^{N-1}\left(-\bar{\psi}^1_{x,y}\psi^2_{x',y}+\bar{\psi}^2_{x',y}\psi^1_{x,y}\right).
 \eb
 \item The identity operator can be represented as a $4N$-fold Grassmann integral
 \begin{eqnarray}\label{res1cst}
 \mathbf{1}_{\mathcal{F}}=\int \mathcal{D}\boldsymbol{\psi}^1_x
 \mathcal{D}\bar{\boldsymbol{\psi}}^1_x\mathcal{D}\boldsymbol{\psi}^2_{x'}
 \mathcal{D}\bar{\boldsymbol{\psi}}^2_{x'}\;
 |\boldsymbol{\psi}^1_{x},\bar{\boldsymbol{\psi}}_{x}^1\rangle
 \langle\boldsymbol{\psi}^2_{x'},\bar{\boldsymbol{\psi}}^2_{x'}|\times\\
 \nonumber\qquad\times \exp\sum_{y=0}^{N-1}\left(
 \bar{\psi}^1_{x,y}\psi^2_{x',y}-\bar{\psi}^2_{x',y}\psi^1_{x,y}\right).
 \end{eqnarray}
 \item The trace of any operator $\mathcal{O}$ can be written as an integral of
 its matrix element in the basis of coherent states with a Gaussian kernel:
 \begin{eqnarray}\label{tracecst}
 \mathrm{Tr}\,\mathcal{O}=
 \int \mathcal{D}\boldsymbol{\psi}^1_x
 \mathcal{D}\bar{\boldsymbol{\psi}}^1_x\mathcal{D}\boldsymbol{\psi}^2_{x'}
 \mathcal{D}\bar{\boldsymbol{\psi}}^2_{x'}\;
 \langle\boldsymbol{\psi}^2_{x'},\bar{\boldsymbol{\psi}}^2_{x'}|\mathcal{O}
 |\boldsymbol{\psi}^1_{x},\bar{\boldsymbol{\psi}}_{x}^1\rangle
 \times
 \\
 \nonumber\qquad\times\exp\sum_{y=0}^{N-1}\left(-\bar{\psi}^1_{x,y}\psi^2_{x',y}+\bar{\psi}^2_{x',y}\psi^1_{x,y}\right).
 \end{eqnarray}
 \item Matrix elements $\langle\boldsymbol{\psi}^2_{x'},\bar{\boldsymbol{\psi}}^2_{x'}|\mathcal{O}
 |\boldsymbol{\psi}^1_{x},\bar{\boldsymbol{\psi}}_{x}^1\rangle$ can be obtained
 by writing $\mathcal{O}$ in normally ordered form, making therein the replacements
 \ben
 a_y\rightarrow \bar{\psi}^1_{x,y},\qquad b_y\rightarrow \psi^1_{x,y},\qquad
 a^{\dag}_y\rightarrow \psi^2_{x',y},\qquad b^{\dag}_y\rightarrow \bar{\psi}^2_{x',y},
 \ebn
 and multiplying the result by (\ref{mecst}).
 \end{itemize}

 Now consider the operator
 \begin{eqnarray}
 \nonumber V_{\alpha}=\; :\exp \sum_{y=0}^{N-1}\Bigl\{b^{\dag}_y(s_y-c_x^{-1}c_y\nabla_{-y})a^{\dag}_y+(
 {c_x^*}^{-1}c_y-1)\left(a^{\dag}_ya_y+b^{\dag}_y b_y\right)+\\
 \label{tm}\qquad +
 a_y(s_y-c_x^{-1}c_y\nabla_y)b_y\Bigr\}:,
 \end{eqnarray}
 where it is understood that
 \be\label{bctwisted}
 \left(\begin{array}{c}a^{\dag}_{y+N} \\ b_{y+N}\end{array}\right)=
 e^{2\pi i \alpha}\left(\begin{array}{c}a^{\dag}_{y} \\ b_{y}\end{array}\right),\quad
  \left(\begin{array}{c}a_{y+N} \\ b^{\dag}_{y+N}\end{array}\right)=
 e^{-2\pi i \alpha}\left(\begin{array}{c}a_{y} \\ b^{\dag}_{y}\end{array}\right).
 \eb
 Rewrite the quantity $Z=\mathrm{Tr}\,V_{\alpha}^M$ using at the first stage (\ref{tracecst}) with
 $x=N-1$, $x'=0$ to calculate the trace, and inserting $M-1$ resolutions of unity (\ref{res1cst})
 (with $x=k-1$ and $x'=k$ between the $k$-th and $(k+1)$-th factor of $V_\alpha$). Then, computing matrix
 elements of $V_{\alpha}$ in the basis of coherent states, the reader may easily check that $Z$
 coincides with the partition function of the Dirac theory described by (\ref{dirop}).

 In fact $V_{\alpha}$ is the transfer matrix characterizing discrete time evolution of twist fields.
 Vertical defects of the action associated to them divide the horizontal axis into intervals.
 The evolution in different intervals is governed by the matrices $V_{\alpha}$ with appropriate
 values of $\alpha$. Twist fields are represented by the operators
 \be\label{twf}
 \mathcal{O}_{\alpha,\alpha'}\left(y^*\right)=\exp 2\pi i \nu \sum_{y'=0}^{y-1}
 \left(-a^{\dag}_{y'}a_{y'}+b^{\dag}_{y'}b_{y'}\right) ,
 \eb
 where $y^*=y-\frac12$. This can be seen by noticing that
 \begin{eqnarray*}
 \fl\exp\biggl\{\delta S_A+\sum_{y'=0}^{N-1}\left(\bar{\psi}^1_{x-1,y'}\psi^2_{x,y'}-
 \bar{\psi}^2_{x,y'}\psi^1_{x-1,y'}\right)\biggr\}=
 \int \mathcal{D}\boldsymbol{\xi}\mathcal{D}\bar{\boldsymbol{\xi}}\;
 \langle\boldsymbol{\xi}^2,\bar{\boldsymbol{\xi}}^2|\mathcal{O}_{\alpha,\alpha'}\left(y^*\right)|
 \boldsymbol{\xi}^1,\bar{\boldsymbol{\xi}}^1\rangle \times \\
 \times \exp\sum_{y'=0}^{N-1}\left(\bar{\psi}^1_{x-1,y'}\xi^2_{y'}-\bar{\xi}^2_{y'}\psi^1_{x-1,y'}
 +\bar{\xi}^1_{y'}\psi^2_{x,y'}-\bar{\psi}^2_{x,y'}\xi^1_{y'}\right),
 \end{eqnarray*}
 and repeating the procedure used above for the computation of $Z$. For example, two-point correlator
 (\ref{2pcf}) can be expressed as
 \ben
 \left\langle\mathcal{O}_{\alpha,\alpha'}(A)\mathcal{O}_{\alpha',\alpha}(B)\right\rangle=
 Z^{-1}\mathrm{Tr}\left(\mathcal{O}_{\alpha,\alpha'}\left(y^*\right)V_{\alpha'}^{x'-x}
 \mathcal{O}_{\alpha',\alpha}\left(y'^*\right)V_{\alpha}^{M-(x'-x)}\right).
 \ebn

 The problem of effective calculation of correlation functions of twist fields therefore reduces to
 the computation of form factors of the operator (\ref{twf})
 between the eigenstates of $V_{\alpha}$ and $V_{\alpha'}$.
 Observe that e.~g. $\mathcal{O}_{\alpha,\alpha'}\left(0^*\right)$ is given by
 the identity operator on $\mathcal{F}$. However, its form factors are
 nontrivial since the corresponding
 bra and ket states diagonalize different transfer matrices.

 \subsection{Transfer matrix diagonalization}\label{sec24}
 Define Fourier transforms of the creation-annihilation operators:
 \begin{eqnarray*}
 \left(\begin{array}{c}
 a^{\dag}_{\theta} \\ b_{\theta}
 \end{array}\right)=\frac{1}{\sqrt{N}}\sum_{y=0}^{N-1}
 \left(\begin{array}{c}
 a^{\dag}_{y} \\ b_{y}
 \end{array}\right) e^{-i\theta y},\\
 \left(\begin{array}{c}
 a_{\theta} \\ b^{\dag}_{\theta}
 \end{array}\right)=\frac{1}{\sqrt{N}}\sum_{y=0}^{N-1}
 \left(\begin{array}{c}
 a_{y} \\ b^{\dag}_{y}
 \end{array}\right) e^{i\theta y},
 \end{eqnarray*}
 where $\theta$ belongs to the set $\boldsymbol{\theta}_{\alpha}=\left\{\frac{2\pi}{N}(k+\alpha)\,|\,k=0,
 \ldots,N-1\right\}$. The only nonvanishing anticommutators are given by
 $\{a_{\theta},a^{\dag}_{\theta'}\}=\{b_{\theta},b^{\dag}_{\theta'}\}=\delta_{\theta\theta'}$.
 The transfer matrix $V_{\alpha}$ is block-diagonal in the Fourier basis,
 \begin{eqnarray*}
 V_{\alpha}=\;:\exp\sum_{\theta\in\boldsymbol{\theta}_{\alpha}}
 \Bigl\{\left(s_y-c_x^{-1}c_ye^{-i\theta}\right)b^{\dag}_{\theta}a^{\dag}_{\theta}
 +({c_x^*}^{-1}c_y-1)\left(a^{\dag}_{\theta}a_{\theta}+
 b^{\dag}_{\theta}b_{\theta}\right)+\\
 \qquad +\left(s_y-c_x^{-1}c_ye^{i\theta}\right)a_{\theta}b_{\theta}\Bigr\}:
 \end{eqnarray*}
 The conjugation of fermions by $V_{\alpha}$ induces linear transformations
 \begin{eqnarray*}
 V_{\alpha}\left(\begin{array}{c}
 a^{\dag}_{\theta} \\ b_{\theta}
 \end{array}\right)V_{\alpha}^{-1}=
 \Lambda(\theta)
 \left(\begin{array}{c}
 a^{\dag}_{\theta} \\ b_{\theta}
 \end{array}\right),\\
  V_{\alpha}\left(\begin{array}{c}
 a_{\theta} \\ b^{\dag}_{\theta}
 \end{array}\right)V_{\alpha}^{-1}=
 \left(\Lambda^{-1}\right)^T(\theta)
 \left(\begin{array}{c}
 a_{\theta} \\ b^{\dag}_{\theta}
 \end{array}\right),
 \end{eqnarray*}
 where $\Lambda(\theta)$ is a Hermitian matrix with unit determinant, explicitly given by
 \ben
 \Lambda(\theta)=\frac{1}{c_y}
 \left(\begin{array}{cc} c_x^*(c_y^2+s_y^2)-2s_x^*s_yc_y\cos\theta &
 -c_x^*s_y+s_x^*c_ye^{i\theta} \\ -c_x^*s_y+s_x^*c_y e^{-i\theta} & c_x^*\end{array}\right).
 \ebn
 It can be brought to the diagonal form,  $ \Lambda(\theta)=U(\theta)
 \left(\begin{array}{cc}e^{-\gamma_{\theta}} & 0 \\
 0 & e^{\gamma_{\theta}}\end{array}\right)U^{\dag}(\theta)$, by a unitary transformation. Here
 $\gamma_{\theta}$ is defined as
 the positive solution of (\ref{gammatheta}),
 and the columns of $U(\theta)=\left(\begin{array}{cr}f_1(\theta) & -\bar{f_2}(\theta) \\
 f_2(\theta) & \bar{f_1}(\theta)\end{array}\right)$ are the eigenvectors of $\Lambda(\theta)$
 normalized so that $|f_1(\theta)|^2+|f_2(\theta)|^2=1$.
 The freedom in the choice of the phase can be used to set
 \be\label{untr}
 \left(\begin{array}{c}
 f_1(\theta) \\ f_2(\theta)
 \end{array}\right)=\frac{1}{2\sqrt{c_y^*}}\left(\begin{array}{cc}
 e^{\Kr_y^*}\; & -e^{-\Kr_y^*} \\ e^{-\Kr_y^*} & \;e^{\Kr_y^*}
 \end{array}\right)\left(\begin{array}{l}
 \chi_{\theta} \\ \chi_{-\theta}
 \end{array}\right),
 \eb
 where
 \be\label{chith}
 \chi_{\theta}=\left[\chi_{-\theta}\right]^{-1}=\left[\frac{(1-\alpha e^{i\theta})(1-\beta e^{-i\theta})}{(1-\beta e^{i\theta})(1-\alpha e^{-i\theta})}\right]^{\frac14}
 \eb
 and $\alpha=\tanh\Kr_x^*\coth\Kr_y$, $\beta=\tanh\Kr_x^*\tanh\Kr_y$. Note that under
 the above conventions one has $\beta<\alpha<1$.
 Root function in (\ref{chith}) is taken on the principal branch.

 A new set of the creation-annihilation operators
 \ben
 \left(\begin{array}{c} c^{\dag}_{\theta} \\ d_{\theta}\end{array}\right)=U^{\dag}(\theta)
 \left(\begin{array}{c} a^{\dag}_{\theta} \\ b_{\theta} \end{array}\right),\qquad
 \left(\begin{array}{c} c_{\theta} \\ d^{\dag}_{\theta}\end{array}\right)=U^T(\theta)
 \left(\begin{array}{c} a_{\theta} \\ b^{\dag}_{\theta} \end{array}\right),
 \ebn
 satisfies canonical anticommutation relations $\{c_{\theta},c^{\dag}_{\theta'}\}=\{d_{\theta},d^{\dag}_{\theta'}\}=\delta_{\theta\theta'}$
 (the other anticommutators being equal to zero)
 and diagonalizes the transfer matrix $V_{\alpha}$:
 \ben
 V_{\alpha}=\left(\frac{c_y}{c_x^*}\right)^N\exp\biggl\{-\sum_{\theta\in\boldsymbol{\theta}_{\alpha}}
 \gamma_{\theta}\left(c^{\dag}_{\theta}c_{\theta}+
 d^{\dag}_{\theta}d_{\theta}-1\right)\biggr\}.
 \ebn
 Introduce the vacua $_{\alpha}\langle vac|$ and $|vac\rangle_{\alpha}$ annihilated
 by all $\{c^{\dag}_{\theta}\}$, $\{d^{\dag}_{\theta}\}$ and $\{c_{\theta}\}$, $\{d_{\theta}\}$, respectively,
 and normalized as $_\alpha\langle vac|vac\rangle_{\alpha}=1$. Left and right eigenvectors
 of $V_{\alpha}$ are then given by the multiparticle Fock states
 \begin{eqnarray}
 \label{lES}
 _{\alpha}\langle\boldsymbol{\theta};\boldsymbol{\phi}|\;\;=&\;_{\alpha}\langle vac|
 c_{\theta_1}\ldots c_{\theta_m}d_{\phi_1}\ldots d_{\phi_n},\\
  \label{rES}
 |\boldsymbol{\theta};\boldsymbol{\phi}\rangle_{\alpha}=&\;
 c^{\dag}_{\theta_1}\ldots c^{\dag}_{\theta_m}d^{\dag}_{\phi_1}\ldots d^{\dag}_{\phi_n}
 |vac\rangle_{\alpha},
 \end{eqnarray}
 and the corresponding eigenvalue is equal to $\left(\frac{c_y}{c_x^*}\right)^N\exp\biggl\{\sum\limits_{\theta\in\boldsymbol{\theta}_{\alpha}}\gamma_{\theta}-
 \sum\limits_{\theta\in\boldsymbol{\theta}\cup\boldsymbol{\phi}}\gamma_{\theta}\biggr\}$.

 The states (\ref{lES})--(\ref{rES}) simultaneously diagonalize the $U(1)$-charge
  and the translation operator,
 \begin{eqnarray*}
 Q=\sum_{\theta\in\boldsymbol{\theta}_{\alpha}}\left(c^{\dag}_{\theta}c_{\theta}-
 d^{\dag}_{\theta}d_{\theta}\right)=
 \sum_{\theta\in\boldsymbol{\theta}_{\alpha}}\left(a^{\dag}_{\theta}a_{\theta}-
 b^{\dag}_{\theta}b_{\theta}\right)=
 \sum_{y=0}^{N-1}\left(a^{\dag}_{y}a_{y}-
 b^{\dag}_{y}b_{y}\right),\\
 \fl \qquad\; T_{\alpha}=\exp\sum_{\theta\in\boldsymbol{\theta}_{\alpha}}{i\theta}\left(-c^{\dag}_{\theta}c_{\theta}
 +d^{\dag}_{\theta}d_{\theta}\right)=
 \; :\exp\sum_{y=0}^{N-1}\left(a^{\dag}_{y}a_{y+1}-a^{\dag}_y a_y+b^{\dag}_{y}b_{y+1}-b^{\dag}_yb_y
 \right):.
 \end{eqnarray*}
 The latter satisfies, e.~g.,
 \ben
 T_{\alpha}\left(\begin{array}{c}a^{\dag}_y \\ b_y\end{array}\right)T_{\alpha}^{-1}=
 \left(\begin{array}{c}a^{\dag}_{y-1} \\ b_{y-1}\end{array}\right),
 \qquad
  T_{\alpha}\left(\begin{array}{c}a_y \\ b^{\dag}_y\end{array}\right)T_{\alpha}^{-1}=
 \left(\begin{array}{c}a_{y-1} \\ b^{\dag}_{y-1}\end{array}\right)
 \ebn
 under boundary conditions (\ref{bctwisted}).

 It is useful to note that the twist field (\ref{twf}) is $U(1)$-neutral and coincides with the identity operator
 twisted by translations of different periodicity:
 \ben
  \mathcal{O}_{\alpha,\alpha'}\left(y^*\right)=
  \left[\mathcal{O}_{\alpha',\alpha}\left(y^*\right)\right]^{-1}=T_{\alpha}^{-y}T_{\alpha'}^{y}.
 \ebn
 This can be seen by comparing linear transformations of fermions induced by the left and right side
 of this relation,
 and using that their action leaves the vector $|vac\rangle$ invariant.
 Form factors of $\mathcal{O}_{\alpha,\alpha'}\left(y^*\right)$ are therefore given, up to simple
 multiplicative factors, by the scalar products of the Fock states (\ref{lES})--(\ref{rES}):
 \ben
 _{\alpha}\langle\boldsymbol{\theta};\boldsymbol{\phi}|
 \mathcal{O}_{\alpha,\alpha'}\left(y^*\right)
 |\boldsymbol{\theta}';\boldsymbol{\phi}'\rangle_{\alpha'}=
 \exp  iy\biggl(\sum_{\theta\in\boldsymbol{\theta}\cup\boldsymbol{\phi}'}\!\!\theta-
 \!\!\sum_{\theta\in\boldsymbol{\phi}\cup\boldsymbol{\theta}'}\!\!\theta\biggr)\;
 {}_{\alpha}\langle\boldsymbol{\theta};\boldsymbol{\phi}
 |\boldsymbol{\theta}';\boldsymbol{\phi}'\rangle_{\alpha'},
 \ebn
 where $\boldsymbol{\theta},\boldsymbol{\phi}\subset\boldsymbol{\theta}_{\alpha}$
 and $\boldsymbol{\theta}',\boldsymbol{\phi}'\subset\boldsymbol{\theta}_{\alpha'}$.
 Such scalar products normalized by the product of the vacua will be denoted by
 \be\label{normmultiff}
 \mathcal{F}_{\alpha,\alpha'}(\boldsymbol{\theta};\boldsymbol{\phi}
 |\boldsymbol{\theta}';\boldsymbol{\phi}')=
 \frac{{}_{\alpha}\langle\boldsymbol{\theta};\boldsymbol{\phi}
 |\mathcal{O}_{\alpha,\alpha'}\left(0^*\right)
 |\boldsymbol{\theta}';\boldsymbol{\phi}'\rangle_{\alpha'}}{
 {}_{\alpha}\langle vac
 |\mathcal{O}_{\alpha,\alpha'}\left(0^*\right)
 |vac\rangle_{\alpha'}\;\,}.
 \eb

 \section{Form factors}
 \subsection{General setting}\label{sec31}
 Let us first recall a few results from \cite{Hystad,IL,Palmer_Hystad}. Consider two sets of $2L$ fermionic creation-annihilation
 operators generating equivalent Fock representations in the same space.
 Denote the corresponding $2^{L}\times2^L$ matrices by
  $\{\psi_i\}$, $\{\psi^{\dag}_i\}$ and $\{\varphi_i\}$, $\{\varphi^{\dag}_i\}$ with
 $i=1,\ldots,L$ and combine them into $L$-columns $\boldsymbol{\psi}$,
 $\boldsymbol{\psi}^{\dag}$, $\boldsymbol{\varphi}$, $\boldsymbol{\varphi}^{\dag}$.
 Suppose there exists a unitary operator $\sigma$ such that
 \ben
 \sigma
 \left(\begin{array}{l}
 \boldsymbol{\varphi}^{\dag} \\ \boldsymbol{\varphi}
 \end{array}\right)
 \sigma^{-1}
 =\left(\begin{array}{cc}
 \mathbf{A} & \mathbf{B} \\ \mathbf{C} & \mathbf{D}
 \end{array}\right)
  \left(\begin{array}{l}
 \boldsymbol{\psi}^{\dag} \\ \boldsymbol{\psi}
 \end{array}\right),
 \ebn
 where $\mathbf{A}$, $\mathbf{B}$, $\mathbf{C}$, $\mathbf{D}$
 are some $L\times L$ matrices. The unitarity of $\sigma$ and canonical
 anticommutation relations imply that
 $\mathbf{B}=\bar{\mathbf{C}}$, $\mathbf{A}=\bar{\mathbf{D}}$ and
 \ben
 \mathbf{D}\mathbf{C}^T+\mathbf{C}\mathbf{D}^T=0,\qquad
 \mathbf{D}\mathbf{D}^{\dag}+ \mathbf{C}\mathbf{C}^{\dag}=\mathbf{1}.
 \ebn

 Suppose that $\mathbf{D}$ is invertible. Then, up to inessential phase factor
 related to the choice of the vacua, one has
 \be\label{vevg}
 _{\boldsymbol{\psi}}\langle vac|\sigma|vac\rangle_{\boldsymbol{\varphi}}=
 |\mathrm{det}\, \mathbf{D}|^{\frac12}.
 \eb
 General matrix elements of $\sigma$ between
 Fock states of different types
 can be expressed as
 \be\label{multiff}
 _{\boldsymbol{\psi}}\langle vac|\psi_{i_1}\ldots\psi_{i_m}\sigma
 \varphi^{\dag}_{j_1}\ldots\varphi^{\dag}_{j_n}|vac\rangle_{\boldsymbol{\varphi}}=
 |\mathrm{det}\, \mathbf{D}|^{\frac12}\cdot \mathrm{Pf}\,R,
 \eb
 \be R=\left(\begin{array}{cc}
 R_{I\times I} & R_{I\times J} \\ R_{J\times I} & R_{J\times J}
 \end{array}\right),
 \eb
 where the entries of the blocks of $(m+n)\times(m+n)$ skew-symmetric matrix
 $R$ are given by the normalized two-particle form factors
 \begin{eqnarray}
 \label{rii}
 \left(R_{I\times I}\right)_{kl}=\left(\mathbf{D}^{-1}\mathbf{C}\right)_{i_k i_l},&
 \qquad k,l=1,\ldots,m,\\
 \label{rij}
 \left(R_{I\times J}\right)_{kl}=-\left(R_{J\times I}\right)_{lk}=\mathbf{D}^{-1}_{i_kj_l},&
 \qquad k=1,\ldots,m,\quad l=1,\ldots,n,\\
 \label{rjj}
 \left(R_{J\times J}\right)_{kl}=\left(\bar{\mathbf{C}}\mathbf{D}^{-1}\right)_{j_kj_l},&
 \qquad k,l=1,\ldots,n.
 \end{eqnarray}

 In the case of interest here, $\sigma$ is the identity operator and $L=2N$.
 The creation-annihilation operators in each set are
 labeled by their $U(1)$-charges and the corresponding momenta. Thus e.~g.
 $\boldsymbol{\psi}$ and $\boldsymbol{\varphi}$ are given by the $2N$-columns
 $\left(\begin{array}{c}\mathbf{c} \\ \mathbf{d} \end{array}\right)$
 built from the operators
 $c_{\theta}$, $d_{\theta}$ with $\theta\in\boldsymbol{\theta}_{\alpha}$ and
 $\theta\in\boldsymbol{\theta}_{\alpha'}$, respectively. $\mathbf{A}$, $\mathbf{B}$,
 $\mathbf{C}$, $\mathbf{D}$ can therefore be seen as  block $2\times 2$ matrices with  block
 entries indexed by $\theta\in\boldsymbol{\theta}_{\alpha'}$, $\theta'\in\boldsymbol{\theta}_{\alpha}$.
 To find their explicit form, note that for $\theta\in\boldsymbol{\theta}_{\alpha'}$ one has
 \begin{eqnarray}
 \label{udagu1}
 \left(\begin{array}{c} c^{\dag}_{\theta} \\ d_{\theta}\end{array}\right)=&\;
 \frac1N\sum_{\theta'\in\boldsymbol{\theta}_{\alpha}} \frac{1-e^{-2\pi i \nu}}{1-e^{i(\theta'-\theta)}}
 \;U^{\dag}(\theta)U(\theta')\left(\begin{array}{c} c^{\dag}_{\theta'} \\ d_{\theta'}\end{array}\right),\\
 \label{udagu2}
 \left(\begin{array}{c} c_{\theta} \\ d^{\dag}_{\theta}\end{array}\right)=&\;
 \frac1N\sum_{\theta'\in\boldsymbol{\theta}_{\alpha}} \frac{1-e^{2\pi i \nu}}{1-e^{i(\theta-\theta')}}
 \;U^{T}(\theta)\bar{U}(\theta')\left(\begin{array}{c} c_{\theta'} \\ d^{\dag}_{\theta'}\end{array}\right),
 \end{eqnarray}
 where $U(\theta)$ is defined by (\ref{untr}).
 Therefore, introducing the notation
 \begin{eqnarray*}
 \Lambda_{\theta,\theta'}=e^{i(\pi\nu+\theta)/2}\delta_{\theta,\theta'},&\qquad \theta,\theta'\in\boldsymbol{\theta}_{\alpha},\qquad\\
  \Lambda'_{\theta,\theta'}=e^{i(\pi\nu-\theta)/2}\delta_{\theta,\theta'},&\qquad
  \theta,\theta'\in\boldsymbol{\theta}_{\alpha'},\qquad\\
 C_{\theta,\theta'}=\frac{\sin\pi\nu}{N}\frac{f_1(\theta){f_2}(\theta')-
 f_2(\theta){f_1}(\theta')}{\sin\frac{\theta'-\theta}{2}},&\qquad
 \theta\in\boldsymbol{\theta}_{\alpha'},\theta'\in\boldsymbol{\theta}_{\alpha},\\
 D_{\theta,\theta'}=\frac{\sin\pi\nu}{N}\frac{f_1(\theta)\bar{f_1}(\theta')+
 f_2(\theta)\bar{f_2}(\theta')}{\sin\frac{\theta'-\theta}{2}},&\qquad
 \theta\in\boldsymbol{\theta}_{\alpha'},\theta'\in\boldsymbol{\theta}_{\alpha},
 \end{eqnarray*}
 we find that
 \ben
 \mathbf{C}=\left(\begin{array}{cc}
 0 & {\Lambda}'C{\Lambda} \\ -\bar{\Lambda}' C \bar{\Lambda} & 0
 \end{array}\right),
 \qquad
  \mathbf{D}=\left(\begin{array}{cc}
 -\Lambda' D \Lambda & 0 \\ 0 &  -\bar{\Lambda}'D\bar{\Lambda}
 \end{array}\right).
 \ebn

 This in turn implies that the vacuum expectation value of twist field and its non-zero
 two-particle form factors are given by
 \be
 \label{ffvev}
 _{\alpha}\langle vac|\mathcal{O}_{\alpha,\alpha'}(0^*)|vac\rangle_{\alpha'}=|\mathrm{det}\,D|,
 \eb
 \begin{eqnarray}
 \label{pffp}
 \mathcal{F}_{\alpha,\alpha'}(\theta;|\theta';)=&\,-
 e^{i(\theta'-\theta-2\pi\nu)/2} D^{-1}_{\theta,\theta'},\\
 \label{mffm}
 \mathcal{F}_{\alpha,\alpha'}(;\theta|;\theta')=&\,-
 e^{i(\theta-\theta'+2\pi\nu)/2} D^{-1}_{\theta,\theta'},\\
 \label{pmff}
 \mathcal{F}_{\alpha,\alpha'}(\theta;\theta'|;)=&\,
 -e^{i(\theta'-\theta)/2}\left(D^{-1}C\right)_{\theta,\theta'},\\
 \label{ffpm}
 \mathcal{F}_{\alpha,\alpha'}(;|\theta;\theta')=&\,
 -e^{i(\theta-\theta')/2} \left(\bar{C}D^{-1}\right)_{\theta,\theta'}.
 \end{eqnarray}
 One also has
 \be
 \label{ffconj}
 \mathcal{F}_{\alpha,\alpha'}(;|\theta;\theta')
 =-\overline{\mathcal{F}_{\alpha',\alpha}(\theta;\theta'|;)},
 \eb
 although this is not immediately obvious from (\ref{pmff})--(\ref{ffpm}).
 This reduces
 our task to the computation of determinant and inverse of $D$ and of the product $D^{-1}C$.

 The relations (\ref{udagu1})--(\ref{udagu2}) imply that the elements of $C$ and $D$ remain invariant
 if one multiplies $U(\theta)$ from the left by a unitary matrix independent of $\theta$. Together
 with (\ref{untr}), this gives
 \begin{eqnarray}
 \label{cchi}
  C_{\theta,\theta'}=&\,\frac{\sin\pi\nu}{2N}\frac{\chi_{\theta}\chi_{-\theta'}-
  \chi_{-\theta}\chi_{\theta'}}{\sin\frac{\theta'-\theta}{2}},\\
 \label{dchi}
  D_{\theta,\theta'}=&\,\frac{\sin\pi\nu}{2N}\frac{\chi_{\theta}\chi_{-\theta'}+
  \chi_{-\theta}\chi_{\theta'}}{\sin\frac{\theta'-\theta}{2}}.
 \end{eqnarray}
 The matrices $C$ and $D$ are therefore very simply related to the ones appearing
 in the Ising model theory, see Lemma~3.3 in \cite{IL}; the main and almost only difference between
 the two cases is the change in the spectrum of quasimomenta.
 We will now follow \cite{IL} to establish elliptic representations for $C$ and $D$,
 which then will be used to calculate $\mathrm{det}\,D$, $D^{-1}$ and $D^{-1}C$.

 \subsection{Elliptic parametrization}\label{sec32}
 Spectral curve (\ref{gammatheta}) is a torus which can be conveniently uniformized
 by the Jacobi elliptic functions of modulus $\displaystyle k=\frac{\sinh2\Kr_x^*}{\sinh2\Kr_y\,}$.
 Let us denote $K=\mathbf{K}(k)$, $K'=\mathbf{K}\bigl(\sqrt{1-k^2}\bigr)$, where $\mathbf{K}(k)$ stands
 for the complete elliptic integral of the first kind.
 Then \cite{IL,Palmer_book} the functions
 \begin{eqnarray}
 \label{zu}
 z(u)=&\,\frac{\mathrm{sn}(u+i\eta)}{\mathrm{sn}(u-i\eta)},\\
 \label{lu}
 \lambda(u)=&\,\left[k\,\mathrm{sn}(u+i\eta)\,\mathrm{sn}(u-i\eta)\right]^{-1},
 \end{eqnarray}
 with $\eta\in\left(-\frac{K'}{2},0\right)$ determined by $\sinh2\Kr_x=i\,\mathrm{sn}\,2i\eta$, satisfy
 the relation (\ref{gammatheta}) written in the form
 \ben
 s_x\left(\lambda+\lambda^{-1}\right)+s_y\left(z+z^{-1}\right)=2c_xc_y.
 \ebn
 The formulas (\ref{zu})--(\ref{lu}) bijectively map the real
 interval $\mathcal{C}_u=\{u\,|\,\mathrm{Re}\,u\in[-K,K),\mathrm{Im}\,u=0\}$ to
 $\mathcal{C}_{\theta}=\left\{(z,\lambda)=(e^{i\theta},e^{\gamma_{\theta}})|\,\theta\in[0,2\pi)\right\}$.
 The inverse image of the point $(e^{i\theta},e^{\gamma_{\theta}})\in\mathcal{C}_{\theta}$ in $\mathcal{C}_{u}$
 will be denoted by $u_{\theta}$. Note that for $\theta\in(0,\pi]$ one has $u_{\theta}=-u_{2\pi-\theta}$.
 It is also useful to define the function
 $\displaystyle x_{\theta}=\frac{\pi u_{\theta}}{2K}$, which
 continuously increases from $-\frac{\pi}{2}$ to $\frac{\pi}{2}$ when $\theta$
 varies from $0$ to $2\pi$.

 Lemma~4.1 in \cite{IL} shows that under the above parametrization matrix elements (\ref{cchi})--(\ref{dchi}) can
 be written as
 \begin{eqnarray}
 \label{cell}
 C_{\theta,\theta'}=&\,\frac{-is_x^* \sin\pi\nu}{N\sqrt{\sinh\gamma_{\theta}\sinh\gamma_{\theta'}}}
 \,\mathrm{cn}(u_{\theta}+u_{\theta'}),\\
 \label{dell}
 D_{\theta,\theta'}=&\,\frac{-s_y \sin\pi\nu}{N\sqrt{\sinh\gamma_{\theta}\sinh\gamma_{\theta'}}}
 \,\frac{\mathrm{dn}(u_{\theta}-u_{\theta'})}{\mathrm{sn}(u_{\theta}-u_{\theta'})}.
 \end{eqnarray}

 We will also need Jacobi theta functions $\vartheta_{1\ldots 4}(z)$ of nome $q=e^{i\pi\tau}$,
 which are related to the elliptic modulus and half-periods by
 \ben
 k=\frac{\vartheta_2^2}{\vartheta_3^2},\qquad 2K=\pi \vartheta_3^2,\qquad 2iK'=\pi\tau\vartheta_3^2,
 \ebn
 where $\vartheta_i=\vartheta_i(0)$ for $i=2,3,4$. The quantities
 $\mathrm{det}\,D$ and $D^{-1}$ can be computed in terms of these
 functions using Frobenius determinant identity,
 while $D^{-1}C$ can be found from a theta functional analog of the Lagrange interpolation formula
 (see Section~5 of \cite{IL} for the details of a similar cumbersome calculation). The result is as follows:
 \begin{eqnarray}
 \nonumber
 \fl\mathrm{det}\,D=
 \frac{\vartheta_3(X_{\alpha'}-X_{\alpha})}{\vartheta_3}
 \left(\frac{s_y\sin\pi\nu}{N}\frac{\vartheta_2\vartheta_4}{\vartheta_3}\right)^N
 (-1)^N\biggl[\;\prod_{\theta\in\boldsymbol{\theta}_{\alpha}}\sinh\gamma_{\theta}
 \prod_{\theta\in\boldsymbol{\theta}_{\alpha'}}\sinh\gamma_{\theta}\biggr]^{-\frac12}\times\\
 \label{detd}
  \times \frac{
 \prod_{\theta,\theta'\in\boldsymbol{\theta}_{\alpha},\theta<\theta'}
 \vartheta_1(x_{\theta'}-x_{\theta})
 \prod_{\theta,\theta'\in\boldsymbol{\theta}_{\alpha'},\theta<\theta'}
 \vartheta_1(x_{\theta}-x_{\theta'})}{
 \prod_{\theta\in\boldsymbol{\theta}_{\alpha},\theta'\in\boldsymbol{\theta}_{\alpha'}}
 \vartheta_1(x_{\theta'}-x_{\theta})}, \\
 \nonumber
 \fl D^{-1}_{\theta,\theta'}=\frac{\vartheta_3(X_{\alpha'}-X_{\alpha}+x_{\theta}-x_{\theta'})}{
 \vartheta_3(X_{\alpha'}-X_{\alpha})\vartheta_1(x_{\theta'}-x_{\theta})}
 \left(\frac{s_y\sin\pi\nu}{N}\frac{\vartheta_2\vartheta_4}{\vartheta_3}\right)^{-1}
 \left[\sinh\gamma_{\theta}\sinh\gamma_{\theta'}\right]^{\frac12} \times\\
 \label{dmone}\times
 \frac{\prod_{\theta''\in\boldsymbol{\theta}_{\alpha'}}\vartheta_1(x_{\theta}-x_{\theta''})
 \prod_{\theta''\in\boldsymbol{\theta}_{\alpha}}\vartheta_1(x_{\theta'}-x_{\theta''})\;\;}{
 \prod_{\theta''\in\boldsymbol{\theta}_{\alpha},\theta''\neq \theta}\vartheta_1(x_{\theta}-x_{\theta''})
 \prod_{\theta''\in\boldsymbol{\theta}_{\alpha'},\theta''\neq \theta'}\vartheta_1(x_{\theta'}-x_{\theta''})
  },\\
 \nonumber
 \fl \left(D^{-1}C\right)_{\theta,\theta'}=-i\frac{\vartheta_2(X_{\alpha'}-X_{\alpha}
 +x_{\theta}+x_{\theta'})}{
 \vartheta_3(X_{\alpha'}-X_{\alpha})}
 \biggl[\frac{\sinh\gamma_{\theta}}{\sinh\gamma_{\theta'}}\biggr]^{\frac12}\times\\
 \label{dmonec}
 \times \frac{\prod_{\theta''\in\boldsymbol{\theta}_{\alpha'}}\vartheta_1(x_{\theta}-x_{\theta''})
 \prod_{\theta''\in\boldsymbol{\theta}_{\alpha},\theta''\neq\theta}\vartheta_4(x_{\theta'}+x_{\theta''})
 }{\prod_{\theta''\in\boldsymbol{\theta}_{\alpha'}}\vartheta_4(x_{\theta'}+x_{\theta''})
 \prod_{\theta''\in\boldsymbol{\theta}_{\alpha},\theta''\neq\theta}\vartheta_1(x_{\theta}-x_{\theta''})
 },
 \end{eqnarray}
 with $X_{\alpha}=\sum_{\theta\in\boldsymbol{\theta}_{\alpha}} x_{\theta}$.
 In the next subsection, the answer (\ref{detd})--(\ref{dmonec}) is rewritten
 in a somewhat different form, which turns out to be more suitable for the
 analysis of the thermodynamic limit and for the computation
 of multiparticle form factors.

 \subsection{VEV and form factors}\label{sec33}
 We illustrate the procedure by transforming the expression for
 $\left(D^{-1}C\right)_{\theta,\theta'}$. First rewrite the second line of (\ref{dmonec})
 as
 \be\label{dmc1}
 \frac{\vartheta_2\vartheta_4}{\vartheta_3}
 \frac{G^-_{\theta}}{G^+_{\theta'}\vartheta_4(x_{\theta}+x_{\theta'})}
 \frac{\prod_{\theta''\in\boldsymbol{\theta}_{\alpha'}}\mathrm{sn}(u_{\theta}-u_{\theta''})}{
 \prod_{\theta''\in\boldsymbol{\theta}_{\alpha},\theta''\neq\theta}\mathrm{sn}(u_{\theta}-u_{\theta''})},
 \eb
 where the functions $G^{\pm}_{\theta}$ are defined by
 \ben\displaystyle G^{\pm}_{\theta}=G^{\mp}_{2\pi-\theta}=
 \frac{\prod_{\theta''\in\boldsymbol{\theta}_{\alpha'}}\vartheta_4(x_{\theta}\pm x_{\theta''})}{
 \prod_{\theta''\in\boldsymbol{\theta}_{\alpha\,}}\vartheta_4(x_{\theta}\pm x_{\theta''})}.
 \ebn
 The last factor in (\ref{dmc1}) can be rewritten using the relation $\displaystyle
 \mathrm{sn}(u_{\theta}-u_{\theta'})=\frac{s_y\sin\frac{\theta-\theta'}{2}}{
 \sinh\frac{\gamma_{\theta}+\gamma_{\theta'}}{2}}$, see e.~g. formula (4.5) in \cite{IL}.
 The resulting products of sine functions can be calculated explicitly: for $\theta\in\boldsymbol{\theta}_{\alpha}$,
 one has
 \begin{eqnarray*}
 2^{N-1}\prod_{\theta''\in\boldsymbol{\theta}_{\alpha'}}\sin\frac{\theta-\theta''}{2}=
 (-1)^{\frac{N\theta}{2\pi}-\alpha+N}\sin\pi\nu,\\
 2^{N-1}\prod_{\theta''\in\boldsymbol{\theta}_{\alpha},\theta''\neq \theta}\sin\frac{\theta-\theta''}{2}=(-1)^{\frac{N\theta}{2\pi}-\alpha+N-1}N.
 \end{eqnarray*}
 The remaining products of sinh's can be combined into the function
 \ben
 H_{\theta}=\frac{\prod_{\theta''\in\boldsymbol{\theta}_{\alpha'}}
 \sinh\frac{\gamma_{\theta}+\gamma_{\theta''}}{2}}{
 \prod_{\theta''\in\boldsymbol{\theta}_{\alpha}}
 \sinh\frac{\gamma_{\theta}+\gamma_{\theta''}}{2}}.
 \ebn
 There exists a simple combination of $G^{\pm}_{\theta}$ and
 $H_{\theta}$  independent of $\theta$, namely
 \be\label{curid}
 \frac{G^{+}_{\theta}G^{-}_{\theta}}{H_{\theta}}=
  \frac{G_{\pi}^2}{H_{\pi}},
 \eb
 with $G_{\pi}=G^{\pm}_{\pi}$. The identity (\ref{curid}) can be proven using the standard addition formulas for theta functions, their relation
 to the elliptic functions and the formula (4.8) in \cite{IL}, relating elliptic and trigonometric
 parametrization.

 Summarizing the above transformations and using (\ref{curid}), we can rewrite the formula (\ref{dmonec}) in the following
 equivalent form:
 \be\label{dmonec2}
 \left(D^{-1}C\right)_{\theta,\theta'}=\frac{is_y\sin\pi\nu\,e^{(\eta_{\theta}+\eta_{\theta'})/2}}{N\sqrt{
 \sinh\gamma_{\theta}\sinh\gamma_{\theta'}}}\frac{\vartheta_2\vartheta_4}{\vartheta_3}
 \frac{\vartheta_2(X_{\alpha'}-X_{\alpha}
 +x_{\theta}+x_{\theta'})}{
 \vartheta_3(X_{\alpha'}-X_{\alpha})\vartheta_4(x_{\theta}+x_{\theta'})},
 \eb
 where the function $\eta_{\theta}$ is given by
 \be\label{etatheta}
 \eta_{\theta}=2\ln\frac{G_{\pi}}{G^+_{\theta}}-\ln H_{\pi}.
 \eb
  Analogous manipulations with (\ref{detd})--(\ref{dmone}) lead to the representations
 \be
 \label{detd2}
 \mathrm{det}\,D=(-1)^N \frac{\vartheta_3(X_{\alpha'}-X_{\alpha})}{\vartheta_3}
  \prod_{\theta\in\boldsymbol{\theta}_{\alpha}}e^{-\eta_{\theta}/4}
 \prod_{\;\theta\in\boldsymbol{\theta}_{\alpha'}}e^{\eta_{\theta}/4},
 \eb
 \be
 \label{dmone2}
 D^{-1}_{\theta,\theta'}=\frac{s_y\sin\pi\nu\,e^{(\eta_{\theta}-
 \eta_{\theta'})/2}}{N\sqrt{
 \sinh\gamma_{\theta}\sinh\gamma_{\theta'}}}\frac{\vartheta_2\vartheta_4}{\vartheta_3}
 \frac{\vartheta_3(X_{\alpha'}-X_{\alpha}+x_{\theta}-x_{\theta'})}{
 \vartheta_3(X_{\alpha'}-X_{\alpha})\vartheta_1(x_{\theta}-x_{\theta'})}.
 \eb
 By (\ref{ffvev}), the formula (\ref{detd2}) gives the vacuum expectation value of twist field.
 Non-zero crossed and non-crossed two-particle form factors are determined by (\ref{pffp}), (\ref{mffm}),
 (\ref{dmone2}) and (\ref{pmff}), (\ref{ffconj}), (\ref{dmonec2}), respectively.

 Multiparticle
 form factors (\ref{normmultiff}) can be obtained from (\ref{multiff})--(\ref{rjj}).
 They do not vanish only if the charge preserving condition $m-m'=n-n'$ is satisfied, where
 $m=\#(\boldsymbol{\theta})$, $n=\#(\boldsymbol{\phi})$, $m'=\#(\boldsymbol{\theta}')$ and $n'=\#(\boldsymbol{\phi}')$. Under this restriction,
 one has
 \begin{eqnarray}
 \fl\nonumber\mathcal{F}_{\alpha,\alpha'}(\boldsymbol{\theta},\boldsymbol{\phi}|
 \boldsymbol{\theta}',\boldsymbol{\phi}')=
 \left(\frac{\vartheta_2
 \vartheta_4}{\vartheta_3}s_y\sin\pi\nu\right)^{m+n'}
 \prod_{\theta\in\boldsymbol{\theta}}\frac{e^{-i(\theta+\pi\nu)/2+\eta_{\theta}/2}}{
 \sqrt{N\sinh\gamma_{\theta}}}
  \prod_{\phi\in\boldsymbol{\phi}}\frac{e^{i(\phi+\pi\nu)/2+\eta_{\phi}/2}}{
 \sqrt{N\sinh\gamma_{\phi}}}\times  \\
 \label{muff}
 \fl\qquad\times\left(-1\right)^{\frac{(m'+n)(m-n'-1)}{2}}
  \prod_{\theta'\in\boldsymbol{\theta}'}\frac{e^{i(\theta'-\pi\nu)/2-\eta_{\theta'}/2}}{
 \sqrt{N\sinh\gamma_{\theta'}}}
  \prod_{\phi'\in\boldsymbol{\phi}'}\frac{e^{-i(\phi'-\pi\nu)/2-\eta_{\phi'}/2}}{
 \sqrt{N\sinh\gamma_{\phi'}}}\cdot\mathrm{det}\,\mathcal{R},
 \end{eqnarray}
 where the $(m+n')\times (m+n')$ matrix $\mathcal{R}$ has $2\times2$ block form,
 $\mathcal{R}=\left(\begin{array}{cc}\mathcal{A} & \mathcal{B} \\ \mathcal{C} & \mathcal{D}
 \end{array}\right)$, and is explicitly given by
 \begin{eqnarray*}
 \mathcal{A}_{jk}=-\frac{i\vartheta_2(X_{\alpha'}-X_{\alpha}
 +x_{\theta_j}+x_{\phi_k})}{
 \vartheta_3(X_{\alpha'}-X_{\alpha})\vartheta_4(x_{\theta_j}+x_{\phi_k})} ,&
 \qquad j=1,\ldots,m,\quad k=1,\ldots, n,\\
 \mathcal{B}_{jk}=-\frac{\vartheta_3(X_{\alpha'}-X_{\alpha}+x_{\theta_j}-x_{\theta'_k})}{
 \vartheta_3(X_{\alpha'}-X_{\alpha})\vartheta_1(x_{\theta_j}-x_{\theta'_k})},&
 \qquad j=1,\ldots,m,\quad k=1,\ldots, m',\\
 \mathcal{C}_{jk}=\quad\;\frac{\vartheta_3(X_{\alpha'}-X_{\alpha}+x_{\phi_k}-x_{\phi'_j})}{
 \vartheta_3(X_{\alpha'}-X_{\alpha})\vartheta_1(x_{\phi_k}-x_{\phi'_j})},&
 \qquad j=1,\ldots,n',\quad k=1,\ldots, n,\\
 \mathcal{D}_{jk}= -\frac{i\vartheta_2(X_{\alpha}-X_{\alpha'}
 +x_{\theta'_k}+x_{\phi'_j})}{
 \vartheta_3(X_{\alpha}-X_{\alpha'})\vartheta_4(x_{\theta'_k}+x_{\phi'_j})} ,&
 \qquad j=1,\ldots,n',\quad k=1,\ldots, m'.
 \end{eqnarray*}

 The determinant of $\mathcal{R}$ in (\ref{muff}) can be evaluated in a closed form using Frobenius
 identity. Given $2L$ indeterminates $z_1,\ldots,z_L$ and $z'_1,\ldots,z'_L$, it expresses
 the determinant of
 the elliptic Cauchy matrix $\Omega$ with elements
 \be\label{ecauchy}
 \Omega_{jk}=\frac{\vartheta_1(z_j-z'_k+r)}{\vartheta_1(r)\vartheta_1(z_j-z'_k)},\qquad
 j,k=1,\ldots,L,\quad r\in\mathbb{C},
 \eb
 in the product form
 \be\label{frobenius}
 \mathrm{det}\,\Omega=\frac{\vartheta_1\Bigl(\sum_{j}^L z_j -\sum_{j}^Lz'_j+r\Bigr)}{\vartheta_1(r)}
 \frac{\prod_{j<k}^L \vartheta_1(z_j-z_k)\vartheta_1(z'_k-z'_j)}{\prod_{j,k}^L\vartheta_1(z_j-z'_k)}.
 \eb
 Setting $L=m+n'$, $r=X_{\alpha'}-X_{\alpha}+\frac{\pi}{2}-\frac{\pi \tau}{2}$ and
 \begin{eqnarray*}
 z_j=&\cases{
 x_{\theta_j}+\frac{\pi\tau}{2} & for $j=1,\ldots,m$,\\
 -x_{\phi'_{j-m}} & for $j=m+1,\ldots,m+n'$,
 }\\
 z'_j=&\cases{
 -x_{\phi_j} & for $j=1,\ldots,n$,\\
 x_{\theta'_{j-n}}+\frac{\pi\tau}{2} & for $j=n+1,\ldots,n+m'$,
 }
 \end{eqnarray*}
 one can check that $\Omega$ coincides with $\mathcal{R}$ up to diagonal matrix factors. It then
 follows from (\ref{frobenius}) that
 \begin{eqnarray}
  \nonumber
 \fl\mathrm{det}\,\mathcal{R}=
 (-1)^{m'(n'+1)}i^{-(m-m')^2}\frac{\vartheta_{p}\Bigl(X_{\alpha'}-X_{\alpha}+
 \sum\limits_{\theta\in\boldsymbol{\theta}}x_{\theta}-
 \sum\limits_{\theta'\in\boldsymbol{\theta'}}x_{\theta'} +
 \sum\limits_{\phi\in\boldsymbol{\phi}}x_{\phi}-
 \sum\limits_{\phi'\in\boldsymbol{\phi}'}x_{\phi'}
 \Bigr)}{\vartheta_3(X_{\alpha'}-X_{\alpha})}\times\\
 \nonumber
 \fl\times
 \frac{\prod\limits_{1\leq j<k\leq m}
 \!\!\vartheta_1(x_{\theta_j}-x_{\theta_k})
 \!\!\prod\limits_{1\leq j<k\leq n}
 \!\!\vartheta_1(x_{\phi_j}-x_{\phi_k})
 \!\!\prod\limits_{1\leq j<k\leq m'}
 \!\!\vartheta_1(x_{\theta'_k}-x_{\theta'_j})
 \!\!\prod\limits_{1\leq j<k\leq n'}
 \!\!\vartheta_1(x_{\phi'_k}-x_{\phi'_j})}{
 \prod\limits_{j=1}^m\prod\limits_{k=1}^{m'}
 \vartheta_1(x_{\theta_j}-x_{\theta'_{k}})
 \prod\limits_{j=1}^{n'}\prod\limits_{k=1}^n
 \vartheta_1(x_{\phi_k}-x_{\phi'_j})
 }\times\\
 \label{detr}
 \times \frac{\prod\limits_{j=1}^m\prod\limits_{k=1}^{n'}
 \vartheta_4(x_{\theta_j}+x_{\phi'_k})
 \prod\limits_{j=1}^n\prod\limits_{k=1}^{m'}
 \vartheta_4(x_{\theta'_k}+x_{\phi_j})}{
 \prod\limits_{j=1}^m\prod\limits_{k=1}^n
 \vartheta_4(x_{\theta_j}+x_{\phi_k})
 \prod\limits_{j=1}^{n'}\prod\limits_{k=1}^{m'}
 \vartheta_4(x_{\theta'_k}+x_{\phi'_j})},
 \end{eqnarray}
 where $p=2$ for $m-m'\in 2\mathbb{Z}+1$ and $p=3$ for $m-m'\in 2\mathbb{Z}$. Together with
 (\ref{muff}), this gives our main result --- a completely explicit factorized formula for any multiparticle
 matrix element of twist field.

 \subsection{Thermodynamic limit}\label{sec34}
 Form factors found above simplify in the thermodynamic limit $N\rightarrow\infty$.
 To explain these simplifications, consider e.~g. the expression (\ref{detd2})
 for the vacuum expectation value (\ref{ffvev}). We need to evaluate the asymptotics
 of (i) $X_{\alpha'}-X_{\alpha}$ and (ii) $\eta_{\theta}$. Both quantities can be written in the
 same form
 \be\label{fsumdiff}
 \sum_{\theta'\in\boldsymbol{\theta}_{\alpha'}}f_{\theta'}-
 \sum_{\theta'\in\boldsymbol{\theta}_{\alpha}}f_{\theta'},
 \eb
 where the function $f_{\theta'}$ is defined for $\theta'\in[0,2\pi]$.
 In the case (i) one has $f_{\theta'}=x_{\theta'}$, while in the case (ii)
 $f_{\theta'}$ is given by
 \be\label{fth}
 f_{\theta'}=2\ln\frac{\vartheta_4(x_{\theta'})}{\vartheta_4(x_{\theta}+x_{\theta'})}-
 \ln \sinh\frac{\gamma_{\pi}+\gamma_{\theta'}}{2}.
 \eb
 The most important distinction between the two situations is that in the first case one has
 $f_{2\pi}-f_{0}=\pi$, whereas (\ref{fth})
 extends to a continuous $2\pi$-periodic function on the real line.

 As $N\rightarrow\infty$, one has the estimate 
 \ben
 f_{\frac{2\pi}{N}(k+\alpha')}-f_{\frac{2\pi}{N}(k+\alpha)}
 \sim \frac{2\pi\nu}{N}f'_{\frac{2\pi k}{N}}
 \ebn 
 with $k=0,\ldots,N-1$. The sum
 $\frac{2\pi}{N}\sum\limits_{k=0}^{N-1}f'_{\frac{2\pi k}{N}}$ transforms into the integral
 $\int\limits_{0}^{2\pi}f'_{\theta}\, d\theta$, which yields
 \be\label{lim12}
 \lim_{N\rightarrow\infty}(X_{\alpha'}-X_{\alpha})=\pi\nu,\qquad
 \lim_{N\rightarrow\infty}\eta_{\theta}=0.
 \eb
 In the same way one shows that
 \be\label{lim3}
 \lim_{N\rightarrow\infty}\biggl(\,\sum_{\theta\in\boldsymbol{\theta}_{\alpha'}}\eta_{\theta}-
 \sum_{\theta\in\boldsymbol{\theta}_{\alpha}}\eta_{\theta}\biggr)=0.
 \eb
 Rigorous mathematical proofs of (\ref{lim12})--(\ref{lim3}) and explicit error estimates
 can be obtained using a kind of Sommerfeld-Watson transform of the sums (\ref{fsumdiff}).

 It then follows from (\ref{ffvev}), (\ref{detd2}) that the thermodynamic limit of
 the vacuum expectation value of twist field is given
 by the simple expression
 \be\label{vevfinal}
 \langle \mathcal{O}_{\nu}\rangle\stackrel{def}{ = }
 \lim_{N\rightarrow\infty}{}_{\alpha}\langle vac|\mathcal{O}_{\alpha,\alpha'}(0^*)|vac\rangle_{\alpha'}=
 \frac{\vartheta_3(\pi\nu|\tau)}{\vartheta_3(0|\tau)},
 \eb
 where the arguments of theta functions and the half-period ratio are now indicated explicitly
 for further convenience. As one could expect on general grounds, the r.h.s. of (\ref{vevfinal}) depends on
 $\alpha$, $\alpha'$ only via the difference $\nu=\alpha'-\alpha$. Note that this relation has the same
  form as the first formula on p.~187 of \cite{Palmer_MF}.

 Particle quasimomenta in the thermodynamic limit uniformly fill the interval $[0,2\pi]$.
 Summation over each of them in the form factor expansions of correlation functions
 transforms into integration:
 $\frac{1}{N}\sum_{\theta}\rightarrow \frac{1}{2\pi}\int_{0}^{2\pi}d\theta$. Although
 this naive procedure is plagued by the appearance of annihilation poles in the crossed
 form factors, it suggests to consider
 \ben
 \mathbf{F}_{\nu}(\boldsymbol{\theta};\boldsymbol{\phi}
 |\boldsymbol{\theta}';\boldsymbol{\phi}')\stackrel{def}{ = }\lim_{N\rightarrow \infty}N^{\frac{m+n+m'+n'}{2}}\mathcal{F}_{\alpha,\alpha'}(\boldsymbol{\theta};\boldsymbol{\phi}
 |\boldsymbol{\theta}';\boldsymbol{\phi}').
 \ebn
 The asymptotics (\ref{lim12}) implies that the limiting two-particle form factors are given by
 \begin{eqnarray}
 \nonumber
 \mathbf{F}_{\nu}(\theta;|\theta';)=\overline{\mathbf{F}_{\nu}(;\theta|;\theta')}=\\
 \label{ffcrth}
 =-\frac{s_y\sin\pi\nu\,
 e^{i(\theta'-\theta-2\pi\nu)/2}}{\sqrt{\sinh\gamma_{\theta}\sinh\gamma_{\theta'}}}
 \frac{\vartheta_2(0|\tau)\vartheta_4(0|\tau)}{\vartheta_3(0|\tau)\vartheta_3(\pi\nu|\tau)}
 \frac{\vartheta_3(x_{\theta}-x_{\theta'}+\pi\nu|\tau)}{\vartheta_1(x_{\theta}-x_{\theta'}|\tau)},\\
 \nonumber
 \mathbf{F}_{\nu}(\theta;\theta'|;)=-\overline{\mathbf{F}_{-\nu}(;|\theta;\theta')}=\\
 \label{ffncth}
 =-\frac{is_y\sin\pi\nu\,
 e^{i(\theta'-\theta)/2}\;\;}{\sqrt{\sinh\gamma_{\theta}\sinh\gamma_{\theta'}}}\;
 \frac{\vartheta_2(0|\tau)\vartheta_4(0|\tau)}{\vartheta_3(0|\tau)\vartheta_3(\pi\nu|\tau)}
 \frac{\vartheta_2(x_{\theta}+x_{\theta'}+\pi\nu|\tau)}{\vartheta_4(x_{\theta}+x_{\theta'}|\tau)}.
 \end{eqnarray}
 The limit of multiparticle form factors can be found in a similar way: it suffices to
 remove the functions
 $\eta_{\theta}$  from (\ref{muff}) and to replace $X_{\alpha'}-X_{\alpha}$ in (\ref{detr}) by $\pi\nu$.

 In the case $\nu=\pm\frac12$, the formulas (\ref{vevfinal})--(\ref{ffncth}) reduce to
 simpler expressions
 \begin{eqnarray}
 \label{vev12}
 \langle \mathcal{O}_{\pm\frac12}\rangle=\left(1-k^2\right)^{\frac14},\\
 \label{ff112}
 \mathbf{F}_{\pm\frac12}(\theta;|\theta';)=\frac{i
 e^{i(\theta'-\theta)/2}}{\sqrt{\sinh\gamma_{\theta}\sinh\gamma_{\theta'}}}
 \frac{\sinh\frac{\gamma_{\theta}+\gamma_{\theta'}}{2}}{\sin\frac{\theta-\theta'}{2}},\\
 \label{ff212}
 \mathbf{F}_{\pm\frac12}(\theta;\theta'|;)=\frac{-is_x^*s_y
 e^{i(\theta'-\theta)/2}}{\sqrt{\sinh\gamma_{\theta}\sinh\gamma_{\theta'}}}
 \frac{\sin\frac{\theta+\theta'}{2}}{\sinh\frac{\gamma_{\theta}+\gamma_{\theta'}}{2}}.
 \end{eqnarray}
 They can be proved using the relations $\vartheta_1\bigl(z+\frac{\pi}{2}\bigr)=\vartheta_2(z)$, $\vartheta_3\bigl(z+\frac{\pi}{2}\bigr)=\vartheta_4(z)$,
 rewriting the theta function ratios in terms of elliptic functions
 and finally passing to the trigonometric parametrization.

 \subsection{Scaling limit}\label{sec35}
 The gap in the energy spectrum closes as $\gamma_0\rightarrow 0$. Since $\gamma_0=2(\Kr_y-\Kr_x^*)$, this
 corresponds to $k\rightarrow 1$
 (or, equivalently, $\tau\rightarrow i0$ or $q\rightarrow1$) in terms of elliptic parameters. Also note that
 \be\label{halfpas}
 -\tau^{-1}=\frac{iK}{K'}= -\frac{i}{\pi}\ln\frac{1-k^2}{16}+o(1)\qquad\mathrm{as}\;k\rightarrow1.
 \eb
 In order to obtain the asymptotics of the vacuum expectation value (\ref{vevfinal}) in
 the vicinity of the critical point, recall the modular transformation
 \be\label{modtr}
 \vartheta_3(z|\tau)=
 (-i\tau)^{-1/2}e^{-\frac{iz^2}{\pi\tau}}\vartheta_3\Bigl(-\frac{z}{\tau}\bigl.\bigr|-\frac{1}{\tau}\Bigr),
 \eb
 and the product formula
 \be\label{thpr}
 \vartheta_3(z|\tau)=\prod_{j=1}^{\infty}(1-q^{2j})(1+q^{2j-1}e^{2iz})(1+q^{2j-1}e^{-2iz}).
 \eb
 Rewrite the theta functions in (\ref{vevfinal}) using (\ref{modtr}). The nome
 of the transformed functions vanishes as $k\rightarrow 1$. The representation
 (\ref{thpr}) therefore implies that
 \ben
 \lim_{\tau\rightarrow i0}\vartheta_3\Bigl(-\frac{\pi\nu}{\tau}\bigl.\bigr|-\frac{1}{\tau}\Bigr)=
 \cases{
 1 &  for $|\nu|<\,\frac12$, \\
 2 &  for $\nu=\pm\frac12$.
 }
 \ebn
 This in turn can be used together with (\ref{halfpas})
 to show that, as $k\rightarrow 1$,
 \be\label{vevas}
 \langle \mathcal{O}_{\nu}\rangle=
 \cases{
 2^{-4\nu^2}\left(1-k^2\right)^{\nu^2}\left[1+o(1)\right] & for $|\nu|<\frac12$,\\
 \qquad\;\left(1-k^2\right)^{\frac14}\,\left[1+o(1)\right] & for $\nu =\pm\frac12$.
 }
 \eb

 The second approximation is in fact exact, cf. (\ref{vev12}).
 As the r.h.s. of (\ref{vevfinal}) is periodic in
 $\nu$ with period~$1$, the critical asymptotics of
 the vacuum expectation value for any $\nu$ can be deduced from (\ref{vevas}).
 Note that the scaling dimension of the twist field for $|\nu|\leq\frac12$ has the expected
 value~$\nu^2$.

 In the vicinity of the critical point, our initial lattice model becomes equivalent to
 a field theory of free massive Dirac fermions. Correlation functions of twist
 fields in this theory are determined by momenta at the scale of inverse
 correlation length (for convenience, we change the domain of definition of lattice momenta
 from $[0,2\pi]$ to $[-\pi,\pi]$).

 More formally, denote $\displaystyle\varepsilon=\frac{1-k^2}{2c_y}$,  set $\theta=\varepsilon \sinh\xi$
 and let $k\rightarrow1$. The dispersion relation (\ref{gammatheta}) then implies that
 $\gamma_{\theta}\rightarrow \varepsilon s_y \cosh \xi$. One also has
 \ben
 -\frac{x_{\theta}-x_{\theta'}}{\tau}\rightarrow
 \frac{i(\xi-\xi')}{2}.
 \ebn
 Normalized scaled two-particle form factors of twist fields are determined by
 \begin{eqnarray*}
 \mathbb{F}_{\nu}(\xi;|\xi';)=&\,\lim_{k\rightarrow 1} \varepsilon\mathbf{F}_{\nu}(\varepsilon\sinh\xi;|\varepsilon\sinh\xi';),\\
 \mathbb{F}_{\nu}(\xi;\xi'|;)=&\,\lim_{k\rightarrow 1} \varepsilon\mathbf{F}_{\nu}(\varepsilon\sinh\xi;-\varepsilon\sinh\xi'|;).
 \end{eqnarray*}
 where the variables $\xi,\xi'$ have the meaning of particle rapidities.

 To compute the corresponding
 limits, one can adopt the same approach as in the above asymptotic analysis of the VEV. Transform
 the theta functions in (\ref{ffcrth})--(\ref{ffncth}) using Jacobi's imaginary transformations, rewrite the result
 using product formulas, and then let $\tau\rightarrow i0$. For $|\nu|<\frac12$, one finds
 \begin{eqnarray*}
 \mathbb{F}_{\nu}(\xi;|\xi';)=&\,
 \frac{\sin\pi\nu}{\sqrt{\cosh\xi\cosh\xi'}}\frac{e^{\nu(\xi'-\xi-i\pi)}}{\sinh\frac{\xi'-\xi}{2}},\\
 \mathbb{F}_{\nu}(\xi;\xi'|;)=&\,
 \frac{\sin\pi\nu}{\sqrt{\cosh\xi\cosh\xi'}}\,\frac{ie^{\nu(\xi'-\xi)}}{\cosh\frac{\xi'-\xi}{2}}\,.
 \end{eqnarray*}
 This reproduces two-particle (and hence all) form factors
 of the continuum twist fields in the massive Dirac theory \cite{Bernard_Leclair,Marino,smj,Truong2},
 which correspond to the exponential fields in the sine-Gordon model at the free-fermion point.

 Likewise, it can be deduced from (\ref{ff112})--(\ref{ff212}) that for $\nu=\pm\frac12$ one has
 \begin{eqnarray*}
 \mathbb{F}_{\pm\frac12}(\xi;|\xi';)=&\,\frac12\left[\mathbb{F}_{\frac12-0}(\xi;|\xi';)+
 \mathbb{F}_{-\frac12+0}(\xi;|\xi';)\right]=\frac{-i\coth\frac{\xi'-\xi}{2}}{\sqrt{\cosh\xi\cosh\xi'}},\\
 \mathbb{F}_{\pm\frac12}(\xi;\xi'|;)=&\,\frac12\left[\mathbb{F}_{\frac12-0}(\xi;\xi'|;)+
 \mathbb{F}_{-\frac12+0}(\xi;\xi'|;)\right]=\frac{i\tanh\frac{\xi'-\xi}{2}}{\sqrt{\cosh\xi\cosh\xi'}}.
 \end{eqnarray*}
 The fact that $\mathbb{F}_{\frac12-0}\neq\mathbb{F}_{-\frac12+0}$ can be
 understood as follows.
 Monodromy conditions for fermion fields in the continuum give a system of integral equations for the
 two-particle form factors \cite{Bernard_Leclair}. This system has a unique admissible solution for $|\nu|<\frac12$, and two solutions for $\nu=\pm\frac12$. Each of the two solutions gives rise to a
 twist field operator. Any linear combination of these operators leads to the required
 fermion branching.
 
  \section{Concluding remarks}
 Finite-lattice form factors in the conventional free-fermion models (triangular Ising lattice,
 XY quantum spin chain, BBS$_2$ model) can be obtained \cite{Iorgov,IL2}
 from those of the Ising spin on the square lattice \cite{PLA,TMF,Iorgov1,Iorgov2,IL}.
 The fields considered in this
 paper are more general; in particular, their scaling dimension continuously depends on a real
 parameter $\nu$. We compute their form factors explicitly in terms of the Jacobi theta functions and show
 that in the scaling limit they reduce to form factors of the exponential fields of the sine-Gordon model
 at the free-fermion point.

 Determinantal form of form factors is also encountered in some of the interacting integrable models,
 see e.~g. Slavnov's formula for scalar products of Bethe states in the spin-$\frac12$ XXZ
 chain \cite{xxz,slavnov}. An intriguing question is therefore if it is possible to go beyond the free-fermion
 point and calculate form factors of twist fields in the integrable lattice regularization of the
 massive Thirring model \cite{Luscher} (related to the eight-vertex statistical model). We believe
 that the present work makes a step in this direction.

 Another challenge is to complete form factor derivation for $\mathbb{Z}_N$-symmetric
 superintegrable chiral Potts quantum chain. It was recently shown that they have Ising form
 up to unknown scalar factors labeled by the pairs of the so-called Onsager sectors \cite{Iorgov4}.
 Putative free-fermion part of the integrable structure of this model is yet to be elucidated.

 \ack
 The authors are grateful to S. Pakuliak and V. Shadura
 for useful discussions. This work was partially supported by the Program of Fundamental
 Research of the Physics and Astronomy Division of NASU,
 by the joint Ukrainian-Russian SFFR-RFBR project F40.2/108, the joint PICS project  of CNRS and NASU,
 and IRSES project ``Random and Integrable Models in Mathematical Physics''.


\Bibliography{50}
 \bibitem{Bernard_Leclair} Bernard D and LeClair A 1994 Differential equations for sine-Gordon
 correlation functions at the free fermion point
 \textit{Nucl. Phys.}~\textbf{B426} 534--558 arXiv:hep-th/9402144v2
 \bibitem{borodin} Borodin A and Deift P 2002 Fredholm
 determinants, Jimbo-Miwa-Ueno tau-functions, and representation
 theory \textit{Comm. Pure Appl. Math.}~\textbf{55} 1160--1230
 	arXiv:math-ph/0111007v1
 \bibitem{PLA} Bugrij A I and Lisovyy O 2003 Spin matrix elements
   in 2D Ising model on the finite lattice
   \textit{Phys. Letts.} \textbf{A319} 390--394 arXiv:0708.3625 [nlin.SI]
 \bibitem{TMF}
    Bugrij A I and Lisovyy O 2004 Correlation function of the two-dimensional Ising model on a finite lattice. II \textit{Theor. Math. Phys.}~\textbf{140} 987--1000 arXiv:0708.3643 [nlin.SI]
 \bibitem{Bugrij_Shadura}
 Bugrij A I and Shadura V N 1999 Asymptotic expression for the correlation function
 of twisted fields in the two-dimensional Dirac model on a lattice
 \textit{Theor. Math. Phys.}~\textbf{121} 1535--1549 arXiv:hep-th/9907040v3
 \bibitem{Doyon} Doyon B 2003 Two-point correlation functions of scaling fields
 in the Dirac theory on the Poincar\'e disk \textit{Nucl.Phys.}~\textbf{B675} 607--630
 	arXiv:hep-th/0304190v2
 \bibitem{Doyon_Fonseca} Doyon B and Fonseca P 2004 Ising field theory on a pseudosphere
 \textit{J. Stat. Mech.}~\textbf{0407} P07002 	arXiv:hep-th/0404136v1
 \bibitem{Doyon_Silk} Doyon B and Silk J 2011 Correlation functions of twist fields
 from Ward identities in the massive Dirac theory \textit{J. Phys.}~\textbf{A44} 295402
 arXiv:1103.2328v1 [hep-th]
  \bibitem{Iorgov1} von Gehlen G, Iorgov N, Pakuliak S, Shadura V and  Tykhyy Yu 2007
 Form-factors in the Baxter-Bazhanov-Stroganov model I: Norms and matrix elements
 \textit{J. Phys.} \textbf{A40} 14117--14138 arXiv:0708.4342 [nlin.SI]
 \bibitem{Iorgov2} von Gehlen G, Iorgov N, Pakuliak S, Shadura V and  Tykhyy Yu 2008
 Form-factors in the Baxter-Bazhanov-Stroganov model II: Ising model on the finite lattice
 \textit{J. Phys.} \textbf{A41} 095003 arXiv:0711.0457 [nlin.SI]
 \bibitem{Hystad}
 Hystad G 2011 Periodic Ising correlations
 \textit{J. Math. Phys.}~\textbf{52} 013302 	arXiv:1011.2223v1 [math-ph]
 \bibitem{Iorgov}
 Iorgov N 2011 Form-factors of the finite quantum XY-chain
 \textit{J. Phys.} \textbf{A44} 335005 arXiv:0912.4466v2 [cond-mat.stat-mech]
 \bibitem{IL}
 Iorgov N and  Lisovyy O 2011 Ising correlations and elliptic determinants 
 \textit{J. Stat. Phys.}~\textbf{143} 33--59 arXiv:1012.2856v2 [math-ph]
 \bibitem{IL2} Iorgov N and Lisovyy O 2011 Finite-lattice form factors in free-fermion models
 \textit{J. Stat. Mech.} P04011 arXiv:1102.2145v2 [cond-mat.stat-mech]
  \bibitem{Iorgov4} Iorgov N, Pakuliak S,   Shadura V, Tykhyy Yu and von Gehlen G 2010
   Spin operator matrix elements in the superintegrable chiral Potts quantum  chain
 \textit{J. Stat. Phys.}~\textbf{139}  743--768 arXiv:0912.5027 [cond-mat.stat-mech]
 \bibitem{Kadanoff}
 Kadanoff L P and Ceva H 1971 Determination of an operator algebra for the two-dimensional Ising model
 \textit{Phys. Rev.}~\textbf{B3} 3918--3939
  \bibitem{xxz} Kitanine N, Maillet J M and Terras V 1998 Form factors of the XXZ
 Heisenberg spin-1/2 finite chain \textit{Nucl. Phys.}~\textbf{B554} 647--678
 	arXiv:math-ph/9807020
 \bibitem{lisovyy_JMP}
 Lisovyy O 2008 On Painlev\'e VI transcendents related to the Dirac operator
   on the hyperbolic disk \textit{J. Math. Phys.}~\textbf{49} 093507 arXiv:0710.5744v1 [math-ph]
 \bibitem{lisovyy_dyson}
 Lisovyy O 2011 Dyson's constant for the hypergeometric kernel 
 \textit{New trends in quantum integrable systems : proceedings of the Infinite Analysis 09} ed B Feigin, M Jimbo and M Okado
  (Singapore: World Scientific) pp 243--267 	arXiv:0910.1914v2 [math-ph]
  \bibitem{lisovyy_peyresq}
  Lisovyy O 2009 Finite-volume correlation functions of monodromy fields
  on the lat\-tice: Toe\-plitz representation \textit{Syst\`emes int\'egrables
 et Th\'eorie des champs quantiques}  ed P Baird, F Helein, J Kouneiher, F Pedit and V Roub\-tsov
 (Paris: Hermann)  pp 170--186.
  \bibitem{LZ}
 Lukyanov S and Zamolodchikov A 1997 Exact expectation values of local fields in the quantum sine-Gordon
 model \textit{Nucl. Phys.} \textbf{B493} 571--587 arXiv:hep-th/9611238
 \bibitem{Luscher}
 L\"uscher M 1976 Dynamical charges in the quantized renormalized massive Thirring model
 \textit{Nucl. Phys.}~\textbf{B117} 475--492
 \bibitem{Marino}
 Marino E C, Schroer B and Swieca J A 1982 Euclidean functional integral approach for
 disorder variables and kinks \textit{Nucl. Phys.}~\textbf{B200} 473--497
 \bibitem{Palmer_book}
 Palmer J 2007 \textit{Planar Ising correlations} (\textit{Prog. Math. Phys.}~\textbf{49} Boston:
 Birkh\"auser)
 \bibitem{Palmer_MF}
 Palmer J 1985 Monodromy fields on $\mathbb{Z}_2$ \textit{Comm. Math. Phys.}~\textbf{102}
 175--206
 \bibitem{Palmer_SD}
 Palmer J 1986 Critical scaling for monodromy fields \textit{Comm. Math. Phys.}~\textbf{104}
  353--385
 \bibitem{PBT} Palmer J, Beatty M and Tracy C A  1994 Tau functions of the Dirac operator on
 the Poincar\'e disk \textit{Comm. Math. Phys.}~\textbf{165} 97--173	arXiv:hep-th/9309017v1
 \bibitem{Palmer_Hystad}
 Palmer J and Hystad G 2010 Spin matrix for the scaled periodic Ising model
 \textit{J. Math. Phys.} \textbf{51} 123301	arXiv:1008.0352v2 [nlin.SI]
  \bibitem{smj}
 Sato M, Miwa T and Jimbo M 1979 Holonomic quantum fields III--IV
 \textit{Publ. RIMS Kyoto Univ.}~\textbf{15} 577--629; \textbf{15} 871--972
 \bibitem{Truong}
 Schroer B and Truong T T 1978 The relativistic quantum fields of the $D=2$ Ising model
 \textit{Phys. Letts.}~\textbf{B72} 371--374
 \bibitem{Truong2}
 Schroer B and Truong T T 1978 The order/disorder quantum field operators associated with the two-dimensional Ising model in the continuum limit \textit{Nucl. Phys.}~\textbf{B144} 80--122
 \bibitem{slavnov}
 Slavnov N 1989 Calculation of scalar products of wave functions and form factors in the framework
 of the algebraic Bethe ansatz \textit{Theor. Math. Phys.}~\textbf{79} 502--508
 \endbib
\end{document}